\documentclass[conference]{IEEEtran}
\IEEEoverridecommandlockouts
\usepackage{cite}
\usepackage{latexsym}
\usepackage{float}
\usepackage{hyperref}
\usepackage{amsmath,amssymb,amsfonts}
\usepackage{algorithmic}
\usepackage{graphicx}
\usepackage{textcomp}
\usepackage{xcolor}
\usepackage{bm}

\def\BibTeX{{\rm B\kern-.05em{\sc i\kern-.025em b}\kern-.08em
    T\kern-.1667em\lower.7ex\hbox{E}\kern-.125emX}}
\begin{document}
\title{MPG-Net: Multi-Prediction Guided Network for Segmentation of Retinal Layers in OCT Images}
\author{\IEEEauthorblockN{Zeyu Fu\textsuperscript{1}, Yang Sun\textsuperscript{2}, Xiangyu Zhang\textsuperscript{3}, Scott Stainton\textsuperscript{4}, Shaun Barney\textsuperscript{4},  Jeffry Hogg\textsuperscript{5}, \\ William Innes\textsuperscript{5} and Satnam Dlay\textsuperscript{3}.}
\IEEEauthorblockA{\textsuperscript{1}Department of Engineering Science, University of Oxford, UK\\
\textsuperscript{2}Big Data Institute, University of Oxford, UK\\
\textsuperscript{3}ISC Research Group, School of Engineering, Newcastle University, UK\\
\textsuperscript{4}Neura Health Group, Newcastle Upon Tyne, UK\\
\textsuperscript{5}Newcastle Eye Centre, Royal Victoria Infirmary, Newcastle Upon Tyne, UK}
\thanks{This work was supported by Medical Research Council Confidence in Concept (grant number
MC\_PC\_17168).}
\thanks{This work was done while Z. Fu and Y. Sun were in Newcastle University. Two authors have equal contributions.}
}

\maketitle
\begin{abstract}
Optical coherence tomography (OCT) is a commonly-used method of extracting high resolution retinal information. Moreover there is an increasing demand for the automated retinal layer segmentation which facilitates the retinal disease diagnosis.
In this paper, we propose a novel multi-prediction guided attention network (MPG-Net) for automated retinal layer segmentation in OCT images.
The proposed method consists of two major steps to strengthen the discriminative power of a U-shape Fully convolutional network (FCN) for reliable automated segmentation.
Firstly, the feature refinement module which adaptively re-weights the feature channels is exploited in the encoder to capture more informative features and discard information in irrelevant regions.
Furthermore, we propose a multi-prediction guided attention mechanism which provides pixel-wise semantic prediction guidance to better recover the segmentation mask at each scale.
This mechanism which transforms the deep supervision to supervised attention is able to guide feature aggregation with more semantic information between intermediate layers.
Experiments on the publicly available Duke OCT dataset confirm the effectiveness of the proposed method as well as an improved performance over other state-of-the-art approaches.
\end{abstract}
\begin{IEEEkeywords}
Optical coherence tomography (OCT), retinal layer segmentation, fully convolutional network (FCN), self attention, semantic prediction.
\end{IEEEkeywords}
\section{Introduction}
Optical coherence tomography (OCT) has become a standard imaging modality in ophthalmic disease diagnosis due to its rapid and non-invasive acquisition of high resolution cross-sectional images of the biological tissue \cite{OCTdia,Multi-dual}.
Retinal OCT scans are particularly being used in the diagnosis of various ocular diseases, and also systemic conditions such as neurodegeneration and cardiovascular disease \cite{OCTreview}. These diseases are often associated with changes in pathological structure within the retina, which yields measurements of retinal layer thickness as biomarkers \cite{Nature}.
Acquiring such biomarkers unavoidably requires manual segmentation in OCT images which is subject to inter-observer variation and is time-consuming. This has motivated many researchers to seek automated retinal layer segmentation methods which are more appropriate for clinical practice.
Established automated segmentation algorithms often use machine learning techniques (e.g. kernel regression \cite{MLOCT} and random forest \cite{MLOCT2}), as well as graph based models \cite{GraphOCT}. However these methods which are not end-to-end learning processes, are limited by manually designed features requiring prior domain knowledge \cite{relay}.

In recent years, numerous approaches have been developed based on convolutional neural networks (CNN), as they have achieved great success in a variety of vision tasks and also become a commonly-used technique in biomedical image segmentation \cite{Multi-dual},\cite{Attention-UNet},\cite{Vnet}. Most of these approaches are inspired from fully convolutional network (FCN) \cite{FCN} or U-Net \cite{Unet}. Inspired from the U-Net, Roy et al. \cite{relay} presented an end-to-end and unified framework for segmenting multiple retinal layers by proposing a U-Net variant with a joint loss function design. To keep accurate topological order of retinal layers, He et al. \cite{Topo} presented a cascaded FCN framework which follows a coarse-to-fine paradigm. Coarse outputs from the first network are further refined by the second network which corrects topological errors. Likewise, Wang et al. \cite{Joint-SC} proposed to jointly detect boundaries and segment retinal layers via a sequential way, in which detected boundaries are employed to enhance the segmentation.
Although cascaded FCNs \cite{Topo},\cite{Joint-SC} have been successful in advancing the segmentation performance, they are complex and require additional computational models. To address this,
attention mechanisms that originally proposed for machine translations \cite{attention-need}, have been recently studied with FCNs for promoting the medical image analysis, such as disease recognition \cite{attn-lesion}, \cite{Multi-dual} and organ segmentation \cite{Attention-UNet}, \cite{Multi-scale}.
Specifically, Oktay et al. \cite{Attention-UNet} proposed an attention U-Net which extends the vanilla U-Net with a novel attention gate for pancreas segmentation.
Rundo et al. \cite{USE-Net} proposed to incorporate self-attention blocks into U-Net for prostate zonal segmentation across from multi-institutional datasets.
In \cite{Multi-dual} and \cite{layer-guided},  enhanced CNN models with multi-level dual attention and layer guided attention respectively were designed for automatic classification of macular diseases.
However, employing attention modules for retinal layer segmentation has not been notably explored.
\begin{figure*}
\centering
\includegraphics[width=1\linewidth]{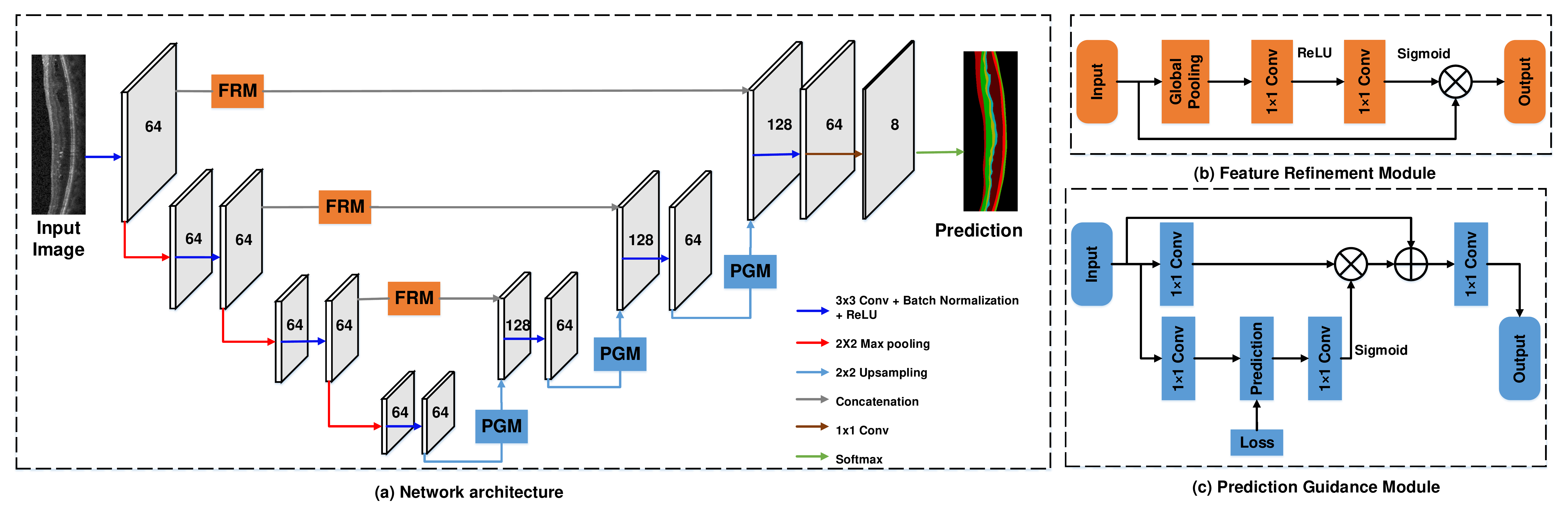}
\caption{Overview of proposed segmentation network. (a) presents the overall network architecture. The digits inside feature maps indicate the number of channels. (b) illustrates the detailed components of feature refinement module. (c) illustrates the detailed components of prediction guidance module. `$\times$' and `$+$' denote multiplication and addition layers respectively.}
\label{overall-system}
\end{figure*}

Therefore, we present a novel multi-prediction guided attention network (MPG-Net) for retinal layer segmentation, as shown in Fig. \ref{overall-system}. The proposed network is based on a
U-shape FCN and is comprised of two attention mechanisms.
Firstly, we harness the feature refinement module (FRM) \cite{SE-Net} originally proposed for natural image analysis, to improve local features in the encoder path. The goal is to transfer more informative features embedded with global context to the enhanced decoder outputs for a better feature fusion, so that the discriminative ability of the network can be improved.
Then we exploit the prediction guidance module (PGM) \cite{SPG-Net} at each decoder output to perform semantic supervision guidance. The PGM was originally developed to propagate useful features across cascaded networks. In contrast, our MPG-Net takes this module a step further, by explicitly developing the supervised attention at each side output to guide the feature aggregation between intermediate layers. Such a design allows the feature maps generated from the decoder path to be enriched with more relevant pixel-wise semantic information from the training targets, so as to better recover the segmentation mask of retinal layers.
Our MPG-Net is evaluated on the Duke OCT dataset \cite{Data}, and the experimental results demonstrate the proposed method produces consistent segmentation improvements and achieves state-of-the-art performance.

The rest of the paper is presented as following: Section \ref{proposed-method} describes the proposed method by two main aspects: feature refinement module and multi-prediction guided attention.  We then explain the experiments, and discuss the results in Section \ref{Experiments}. The final section concludes the paper with some future work.

\section{Proposed Method}
\label{proposed-method}
Fig. \ref{overall-system} illustrates an overview of proposed segmentation network, which contains a modified U-shape FCN \cite{Unet} as the backbone in the overall network, and also embeds with the FRM and PGM.
Our end-to-end MPG-Net consists of four contractive encoders and three expansive decoders, which is similar to the original design of U-Net \cite{Unet}.
For each encoder, the inputs are performed with a $3\times3$ convolutional layer, a batch normalization and a rectified linear unit (ReLU) \cite{relu} activation function for the feature extraction. The max pooling layer is used as a downsampling module in each encoder path. Before each skip connection which concatenates extracted features maps to the decoders, the FRM is applied to integrate more global context information with local features adaptively.
Each decoder conducts the same method as in encoders to generate feature maps, and they are further processed in the PGM to improve their representative ability via semantic prediction guidance at each side output.
The decoder path is also equipped with an upsampling layer to recover feature resolution using bilinear interpolation.

Finally, outputs from the last decoder are fed into a pixel-wise classifier which employs a $1\times1$ convolutional layer with channel-wise Softmax activation to produce the final prediction. For multi-class semantic segmentation, the output of our network includes $K$ channels, where $K$ denotes the number of classes including seven different retinal layers and the background ($K = 8$).

\subsection{Feature Refinement Module}
It is known that the success of U-Net \cite{Unet} has significantly advanced the medical image segmentation, due to its encoder-decoder architecture with skip connections. However, the standard encoder design may not capture representative features sufficiently without fully considering long-range contextual relationships \cite{Multi-scale}.
Recent literature \cite{SE-Net} to mitigate the aforementioned issue demonstrates that reweighting the feature channels can highlight informative features and discard irrelevant ones in the image classification task. Therefore, we follow this trend and then exploit the designed FRM \cite{SE-Net} to deliver more salient features incorporated with global context to the decoder for the retinal layer segmentation.

The detailed structure of the FRM is depicted in Fig. \ref{overall-system} (b). Given the feature map $\mathbf{F}\in \mathbb{R}^{H\times W\times C}$ after the encoder, a global average pooling layer \cite{global-pooling} which provides the global receptive field is firstly applied to capture the global context information. This yields a context vector $\mathbf{g} \in \mathbb{R}^{C}$, and its $c$-th element can be given as \cite{SE-Net},
\begin{equation}
  g_{c} = \frac{1}{H\times W}\sum^{H}_{i=1}\sum^{W}_{j=1}\mathbf{f}_{c}(i,j)
\end{equation}
where $\mathbf{f}_{c}$ denotes a single channel feature map with the spatial size of $H\times W$, and $C$ denotes the number of channels.
Then two $1\times1$ convolutional layers with ReLU and Sigmoid activation functions respectively are applied on the vector $\mathbf{g}$ to generate a channel-wise attention map $\mathbf{S}_{e}\in \mathbb{R}^{1\times 1\times C}$, which rescales the input feature map $\mathbf{F}$ into the output $\tilde{\mathbf{F}}$ via channel-wise multiplication.
By employing the FRM in our network, local features in the encoder path are refined with their corresponding global context and thereby boosting their representative power.

\subsection{Multi-Prediction Guided Attention}
In the decoder path, to better recover the feature resolution at each scale with more semantic information, existing approaches \cite{Attention-UNet},\cite{Multi-scale}, \cite{M-Net} tend to apply deep supervision \cite{deep-supervision} to improve the semantic discriminability between intermediate feature maps. Motivated by this, we propose to transform the deep supervision into multi-prediction guided attention by taking advantage of the PGM \cite{SPG-Net} which provides semantic prediction guidance, as shown in Fig. \ref{overall-system} (c).

Specifically, let $\mathbf{F}_{d}\in \mathbb{R}^{H\times W\times C}$ denote a decoder output, and it is firstly fed to a $1\times1$ convolutional layer to compute per-class logits $\mathbf{F}_{k}\in \mathbb{R}^{H\times W\times K}$ \cite{SPG-Net}. Followed by a $1\times1$ convolutional layer with Sigmoid activation function, a prediction guided attention map $\mathbf{S}_{d}\in \mathbb{R}^{H\times W\times C}$ can be then generated from $\mathbf{F}_{k}$. On the other hand, the decoder output $\mathbf{F}_{d}$ is also processed with a $1\times1$ convolutional layer to produce a transformed decoder feature map, which is then multiplied with the attention map $\mathbf{S}_{d}$ to obtain a reweighted feature map.
Finally, we add this reweighted feature map with the decoder output $\mathbf{F}_{d}$, as well as applying another $1\times1$ convolution layer to compute the input feature to the next decoder.
This prediction guidance is learned by optimizing the loss between semantic prediction and the corresponding scale of the ground truth.
The advantage of our proposed multi-prediction guided attention is that it can guide intermediate feature maps to be semantically discriminant at pixel level rather than the image level in deep supervision \cite{deep-supervision}.

For the network training, we utilize a primary loss function to supervise the final prediction of the entire MPG-Net, and also apply three auxiliary loss functions to jointly supervise the outputs from intermediate layers.
All the loss functions have the same design as illustrated in (\ref{customized-loss}), which consists of a multi-class cross-entropy loss in (\ref{customized-loss1}) and a dice loss in (\ref{customized-loss2}) which calculates the spatial overlapping between the predictions and the corresponding ground truth \cite{relay} \cite{DiceLoss}. Given a pixel $x$ in the image $\mathbf{X}$, our loss function can be formulated as follows,
\begin{equation}\label{customized-loss}
L_{seg}={\alpha}L_{CE}+{\beta}L_{Dice}
\end{equation}
\begin{equation}\label{customized-loss1}
L_{CE} = -\sum_{x \in \mathbf{X}}y_{k}(x)\log(p_{k}(x))
\end{equation}
\begin{equation}\label{customized-loss2}
L_{Dice} = 1-\frac{2\sum_{x\in \mathbf{X}}y_{k}(x)p_{k}(x)}{\sum_{x\in \mathbf{X}}p^2_k(x)+\sum_{x\in \mathbf{X}}y^2_k(x)}
\end{equation}
where $p_{k}(x)\in [0,1]$  denotes the predicted probability of pixel $x$ belonging to class $k\in K$, and $y_k(x) \in \{0,1\}$ represents the binary ground truth label of pixel $x$ for class $k$.
\begin{figure}[htbp!]
\centering
\includegraphics[width=1\linewidth]{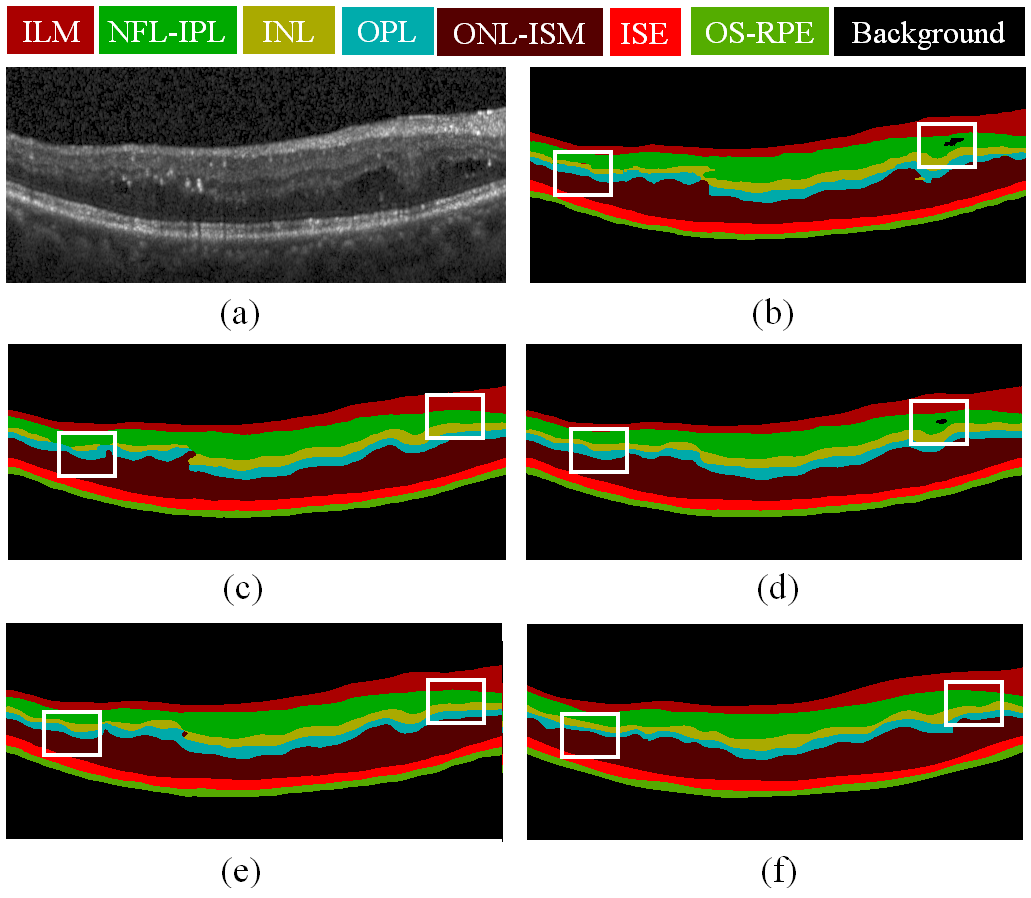}
\caption{Qualitative comparison for the retinal layer segmentation results from the ablation setting. (a) Original input image.  (b) Baseline method. (c) Baseline with FRM successfully removes the irrelevant background regions on the right, but fails to rectify the false positives among retinal layers. (d) Baseline with MPGA well classifies among the foreground layers on the left, while it is sensitive to the background.
(e) Baseline with FRM and MPGA performs the best. (f) Ground Truth (expert 1). Different colors corresponding to retinal layers are shown on the top (better viewed in color version).}
\label{visual-results}
\end{figure}
The hyper parameters of $\alpha$ and $\beta$ are used to balance the weight between the $L_{CE}$ and $L_{Dice}$. Finally, we follow the design of \cite{M-Net} and combine the primary and auxiliary loss to formulate the following joint loss function,
\begin{equation}
L = \sum_{n=1}^{N} L_{seg}^{(n)}
\end{equation}
where $N$ is the number of output layers.

\section{Experiments and Results}
\label{Experiments}
This section firstly presents the well-established dataset, then explains the detailed settings of our MPG-Net and how it implements for the retinal layer segmentation. Next, an ablation study is performed to analyze the effects of each module in the proposed network. Finally, we compare our MPG-Net with other state-of-the-art approaches and discuss their results.
\begin{table*}[htbp!]
\caption{Quantitative results of retinal layer segmentation evaluated on the Duke OCT dataset \cite{Data}, \textbf{Bold} indicates the best results.}
\begin{tabular}{|c|c|c|c|c|c|c|c|c|}
\hline
Algorithm &  Background & ILM & NFL-IPL & INL & OPL & ONL-ISM &ISE &OS-RPE\\
\hline
\hline

Attention U-Net \cite{Attention-UNet} &  0.991& 0.871 & 0.904 &0.777&0.762 &0.927 &0.889&0.850\\

Chiu et al. \cite{MLOCT} &  NA & 0.851&0.892&0.754&0.740&0.930&0.871&0.820\\

Long  et al. \cite{FCN}&  0.971 & 0.812&  0.843& 0.717& 0.711& 0.883& 0.885& \textbf{0.861}\\

Expert 2 \cite{Data} &  NA & 0.860 &  0.903& \textbf{0.797}& 0.747& \textbf{0.941} &0.862 & 0.829\\

Two-stage \cite{Two-stage} &  0.985 & 0.837& 0.898& 0.781& 0.763& 0.931& 0.887 &0.824\\
Bi-decision \cite{Joint-SC} &  NA & 0.860 &  0.900&0.780 & \textbf{0.780}& 0.940 &0.862 & 0.829\\
\hline
\hline
Baseline & 0.979& 0.837 &0.896&0.770 &0.762 &0.910 &0.889 & 0.780 \\
Baseline $+$ FRM &  0.991 & 0.878& 0.908& 0.779& 0.764& 0.928& 0.891 &0.848\\

Baseline $+$ MPGA &  \textbf{0.992} & 0.880& 0.909& 0.786& 0.771& 0.932& 0.895 &0.854\\

Full model (MPG-Net) &  \textbf{0.992} & \textbf{0.882}& \textbf{0.910}& 0.794& 0.779& 0.935& \textbf{0.897} &0.857\\
\hline
\end{tabular}
\centering
\label{quantative-results}
\end{table*}
\subsection{Dataset and Evaluation Metrics}
The OCT data used to evaluate the performance of the proposed method is taken from the publicly available Duke OCT dataset \cite{Data}. This dataset contains 110 OCT images from 10 Diabetic Macular Edema (DME) patients annotated by two experts, where each patient is assigned with 11 OCT scans. To achieve a fair comparison,  we follow the default setting in \cite{Joint-SC} to split these labelled images into a training set (patients 1-5) and a testing set (patients 6-10). For training/testing the network, annotations from the expert 1 are used as ground truth, while the expert 2 is used for comparison. In addition, the F1 score which is a commonly-used metric in biomedical image segmentation is utilized as the performance measure to evaluate the segmentation performance.

\subsection{Implementation Details}
Our MPG-Net is implemented using Keras framework with Tensorflow backend.
Input images are resized into  $216\times496$ for the network implementation.
The Adam optimizer \cite{Adam} is used to optimize the deep model, with the mini batch size of $2$. Other settings to train the model are presented as follows: the number of epochs is $100$, a decaying learning rate is used with a starting point of $0.01$, and a weight decay rate of $10^{-4}$ ($L2$ regularization) is applied to regularize the network.
Parameters of $\alpha$ and $\beta$ used in (\ref{customized-loss}) are empirically chosen as $1$ and $0.5$ respectively.
All the experimental results were achieved by means of a work station with an Intel i7 CPU, a GeForce RTX 2080Ti GPU, and 64GB of RAM.

\subsection{Ablation Study}

In this section, we present an ablation study to individually examine the effect of the feature refinement module (FRM) and multi-prediction guided attention (MPGA) on the segmentation performance. The baseline method is denoted by disabling both attention modules in our MPG-Net. Models compared in this study were trained with the same parameter settings for evaluation. For each model, we report the F1 scores to check the segmentation performance for each layer, including internal limiting membrane (ILM), nerve fiber layer-inner plexiform layer (NFL-IPL), inner nuclear layer (INL), outer plexiform layer (OPL), outer nuclear layer-inner segment myeloid (ONL-ISM), inner segment ellipsoid (ISE) and outer segment to retinal pigment epithelium (OS-RPE) \cite{seven}.

The results for this ablation experiment are reported in the lower part of Table \ref{quantative-results}.
As we can see from the first two rows, incorporating the FRM into the baseline architecture achieves noticeable improvements for all the segmented layers, especially in OS-RPE layer with $6.8\%$. Similarly, when integrating the MPGA with the baseline model, the F1 scores of Baseline are generally increased between $0.6\%-7.4\%$ on average.
More importantly, the proposed MPG-Net which fuses both attention modules continues to improve the baseline performance and reports the best scores among compared models.
On the other hand, Fig. \ref{visual-results} visually demonstrates how each proposed module contributes to mitigating the misalignment between the prediction and ground truth.
Fig. \ref{visual-results} (c) shows that using FRM is advantageous, as it emphasizes the feature saliency of foreground objects and meanwhile suppresses the less useful background areas. As for Fig. \ref{visual-results} (d), it demonstrates that the MPGA is able to strengthen the semantic discriminability among the retinal layers.
Besides, the result as shown in Fig. \ref{visual-results} (e) confirms the combined design can better exploit the complementary benefits from both attention models.

\subsection{Comparison with State-of-the-art}
The proposed method is then evaluated on the same dataset to compare its performance with the Expert 2 \cite{Data}  as well as other state-of-the-art methods, including Attention U-Net \cite{Attention-UNet}, Chiu et al. \cite{MLOCT}, Long et al. \cite{FCN}, Two-stage \cite{Two-stage} and Bi-decision \cite{Joint-SC}.
As we can see from Table \ref{quantative-results}, the proposed method (MPG-Net) reports the best scores in segmenting layers of ILM, NFL-IPL, ISE, and background. Meanwhile, it performs the second best in terms of OPL, INL, and OS-RPE, where the OPL layer is considered as the most challenging layer to be segmented \cite{relay}.
Specifically, our deep model favorably surpasses the manual expert 2 in several layers,
demonstrating an alternative to tedious manual segmentation methods.
Comparing to the cascaded network in \cite{Joint-SC},
the proposed method performs better in most layers without requiring additional computational models or any post-processing steps. Besides, MPG-Net outperforms the existing attention approach \cite{Attention-UNet} in all layers, indicating the proposed method is more capable to learn and highlight informative features and thereby eliminating uncertainties.
Evaluations above suggest that the use of our proposed attention mechanism can enhance the generalization ability of encoder-decoder architecture, resulting in more promising segmentation performance.

\section{Conclusions and Future Work}
\label{Conclusions}
In this paper, we proposed a novel attention based network (MPG-Net) for retinal layer segmentation in OCT images. The proposed attention model was developed to improve the generalization ability of the standard encoder-decoder architecture by combining the multi-prediction guided attention and feature refinement at multiple scales. The feature refinement module (FRM) was incorporated in each encoder to highlight salient features related to foreground objects by adaptively reweighting local features with global context information.
Moreover, the multi-prediction guided attention was proposed to increase the feature discriminability among retinal layers via the semantic supervision guidance.
Ablation experiments as well as comparisons with state-of-the-art methods confirmed the effectiveness of the proposed method.
Future work will concentrate on exploring more structural information from the OCT scans, in order to further increase the segmentation performance.
\bibliographystyle{IEEEtran}
\bibliography{OCT}

\begin{thebibliography}{10}
\providecommand{\url}[1]{#1}
\csname url@samestyle\endcsname
\providecommand{\newblock}{\relax}
\providecommand{\bibinfo}[2]{#2}
\providecommand{\BIBentrySTDinterwordspacing}{\spaceskip=0pt\relax}
\providecommand{\BIBentryALTinterwordstretchfactor}{4}
\providecommand{\BIBentryALTinterwordspacing}{\spaceskip=\fontdimen2\font plus
\BIBentryALTinterwordstretchfactor\fontdimen3\font minus
  \fontdimen4\font\relax}
\providecommand{\BIBforeignlanguage}[2]{{%
\expandafter\ifx\csname l@#1\endcsname\relax
\typeout{** WARNING: IEEEtran.bst: No hyphenation pattern has been}%
\typeout{** loaded for the language `#1'. Using the pattern for}%
\typeout{** the default language instead.}%
\else
\language=\csname l@#1\endcsname
\fi
#2}}
\providecommand{\BIBdecl}{\relax}
\BIBdecl

\bibitem{OCTdia}
F.~A. Medeiros, L.~M. Zangwill, L.~M. Alencar, C.~Bowd, P.~A. Sample, J.~R.
  Susanna, and R.~N. Weinreb, ``Detection of glaucoma progression with stratus
  {OCT} retinal nerve fiber layer, optic nerve head, and macular thickness
  measurements,'' \emph{Investigative Ophthalmology $\&$ Visual Science},
  vol.~50, no.~12, pp. 5741--5748, 2009.

\bibitem{Multi-dual}
S.~S. {Mishra}, B.~{Mandal}, and N.~B. {Puhan}, ``Multi-level dual-attention
  based {CNN} for macular optical coherence tomography classification,''
  \emph{IEEE Signal Processing Letters}, vol.~26, no.~12, pp. 1793--1797, 2019.

\bibitem{OCTreview}
M.~R. Hee, J.~A. Izatt, E.~A. Swanson, D.~Huang, J.~S. Schuman, C.~P. Lin,
  C.~A. Puliafito, and J.~G. Fujimoto, ``Optical coherence tomography of the
  human retina,'' \emph{Arch Ophthalmol}, vol. 113, no.~3, pp. 325--332, 1995.

\bibitem{Nature}
J.~D. Fauw, J.~R. Ledsam, B.~Paredes, S.~Nikolov, N.~Tomasev, S.~Blackwell,
  H.~Askham, and et~al., ``Clinically applicable deep learning for diagnosis
  and referral in retinal disease,'' \emph{Nature Medicine}, vol.~24, no.~9,
  pp. 1342--1350, 2018.

\bibitem{MLOCT}
S.~J. Chiu, M.~J. Allingham, P.~S. Mettu, S.~W. Cousins, J.~A. Izatt, and
  S.~Farsiu, ``Kernel regression based segmentation of optical coherence
  tomography images with diabetic macular edema,'' \emph{Biomedical Optics
  Express}, vol.~6, no.~4, pp. 1172--1194, 2015.

\bibitem{MLOCT2}
S.~P.~K. Karri, D.~Chakraborthi, and J.~Chatterjee, ``Learning layer-specific
  edges for segmenting retinal layers with large deformations,''
  \emph{Biomedical Optics Express}, vol.~7, no.~7, pp. 2888--2901, 2016.

\bibitem{GraphOCT}
P.~A. Dufour, L.~Ceklic, H.~Abdillahi, S.~Schröder, S.~D. Dzanet,
  U.~W.-Schnurrbusch, and J.~Kowal, ``Graph-based multi-surface segmentation of
  {OCT} data using trained hard and soft constraints,'' \emph{IEEE Transactions
  on Medical Imaging}, vol.~32, no.~3, pp. 531--543, 2013.

\bibitem{relay}
A.~G. Roy, S.~Conjeti, S.~P.~K. Karri, D.~Sheet, A.~Katouzian, C.~Wachinger,
  and N.~Navab, ``Relaynet: retinal layer and fluid segmentation of macular
  optical coherence tomography using fully convolutional networks,''
  \emph{Biomedical Optics Express}, vol.~8, no.~8, pp. 3627--3642, 2017.

\bibitem{Attention-UNet}
O.~Oktay, J.~Schlemper, L.~L. Folgoc, M.~Lee, M.~Heinrich, K.~Misawa, K.~Mori,
  S.~McDonagh, N.~Y. Hammerla, B.~Kainz, B.~Glocker, and D.~Rueckert,
  ``Attention {U-Net}: Learning where to look for the pancreas,'' in
  \emph{International Conference on Medical Imaging with Deep Learning (MIDL)},
  2018, pp. 1--10.

\bibitem{Vnet}
F.~Milletari, N.~Navab, and S.-A. Ahmadi, ``{V-Net}: fully convolutional neural
  networks for volumetric medical image segmentation,'' in \emph{Proc. of
  International Conference on 3D Vision}, 2016, pp. 565--571.

\bibitem{FCN}
J.~Long, E.~Shelhamer, and T.~Darrell, ``Fully convolutional networks for
  semantic segmentation,'' in \emph{Proc. of Computer Vision and Pattern
  Recognition (CVPR)}.\hskip 1em plus 0.5em minus 0.4em\relax IEEE, 2015, pp.
  3431--3440.

\bibitem{Unet}
O.~Ronneberger, P.~Fischer, and T.~Brox, ``{U-Net}: Convolutional networks for
  biomedical image segmentation,'' in \emph{Proc. of Medical Image Computing
  and Computer-Assisted Intervention (MICCAI)}.\hskip 1em plus 0.5em minus
  0.4em\relax Springer, 2015, pp. 234--241.

\bibitem{Topo}
Y.~He, A.~Carass, Y.~Yun, C.~Zhao, B.~M. Jedynak, S.~D. Solomon, S.~Saidha,
  P.~A. Calabresi, and J.~L. Prince, ``Towards topological correct segmentation
  of macular {OCT} from cascaded {FCNs},'' in \emph{Proc. of Medical Image
  Computing and Computer-Assisted Intervention (MICCAI)}, 2017, pp. 202--209.

\bibitem{Joint-SC}
J.~Wang, Z.~Wang, F.~Li, G.~Qu, Y.~Qiao, H.~Lv, and X.~Zhang, ``Joint retina
  segmentation and classification for early glaucoma diagnosis,''
  \emph{Biomedical Optics Express}, vol.~10, no.~5, pp. 2639--2656, 2019.

\bibitem{attention-need}
A.~Vaswani, N.~Shazeer, N.~Parmar, J.~Uszkoreit, L.~Jones, A.~N. Gomez, L.~u.
  Kaiser, and I.~Polosukhin, ``Attention is all you need,'' in \emph{Advances
  in Neural Information Processing Systems}, 2017, pp. 5998--6008.

\bibitem{attn-lesion}
L.~{Fang}, C.~{Wang}, S.~{Li}, H.~{Rabbani}, X.~{Chen}, and Z.~{Liu},
  ``Attention to lesion: Lesion-aware convolutional neural network for retinal
  optical coherence tomography image classification,'' \emph{IEEE Transactions
  on Medical Imaging}, vol.~38, no.~8, pp. 1959--1970, 2019.

\bibitem{Multi-scale}
A.~Sinha and J.~Dolz, ``Multi-scale guided attention for medical image
  segmentation,'' \emph{arXiv:1906.02849 [cs.CV]}, pp. 1--11, 2019.

\bibitem{USE-Net}
L.~Rundo, C.~Han, Y.~Nagano, J.~Zhang, R.~Hataya, C.~Militello, A.~Tangherloni,
  M.~S. Nobile, C.~Ferretti, D.~Besozzi, M.~C. Gilardi, S.~Vitabile, G.~Mauri,
  H.~Nakayama, and P.~Cazzaniga, ``{USE-Net}: Incorporating
  squeeze-and-excitation blocks into {U-Net} for prostate zonal segmentation of
  multi-institutional {MRI} datasets,'' \emph{Neurocomputing}, vol. 365, pp. 31
  -- 43, 2019.

\bibitem{layer-guided}
L.~{Huang}, X.~{He}, L.~{Fang}, H.~{Rabbani}, and X.~{Chen}, ``Automatic
  classification of retinal optical coherence tomography images with layer
  guided convolutional neural network,'' \emph{IEEE Signal Processing Letters},
  vol.~26, no.~7, pp. 1026--1030, 2019.

\bibitem{SE-Net}
J.~{Hu}, L.~{Shen}, and G.~{Sun}, ``Squeeze-and-excitation networks,'' in
  \emph{IEEE Conference on Computer Vision and Pattern Recognition (CVPR)},
  2018, pp. 7132--7141.

\bibitem{SPG-Net}
B.~Cheng, L.-C. Chen, Y.~Wei, Y.~Zhu, Z.~Huang, J.~Xiong, T.~Huang, W.-M. Hwu,
  and H.~Shi, ``{SPGNet}: Semantic prediction guidance for scene parsing,'' in
  \emph{Proc. of IEEE International Conference on Computer Vision (ICCV)},
  2019, pp. 5218--5228.

\bibitem{Data}
P.~P. Srinivasan, L.~A. Kim, P.~S. Mettu, S.~W. Cousins, G.~M. Comer, J.~A.
  Izatt, and S.~Farsiu, ``Fully automated detection of diabetic macular edema
  and dry age-related macular degeneration from optical coherence tomography
  images,'' \emph{Biomedical Optics Express}, vol.~5, no.~10, pp. 3568--3577,
  2014.

\bibitem{relu}
X.~Glorot, A.~Bordes, and Y.~Bengio, ``Deep sparse rectifier neural networks,''
  in \emph{Proc. of Artificial Intelligence and Statistics (AISTATS)}, 2011,
  pp. 315--323.

\bibitem{global-pooling}
W.~Liu, A.~Rabinovich, and A.~C. Berg, ``Parsenet: Looking wider to see
  better,'' \emph{arXiv:1506.04579 [cs.CV]}, pp. 1--11, 2015.

\bibitem{M-Net}
H.~{Fu}, J.~{Cheng}, Y.~{Xu}, D.~W.~K. {Wong}, J.~{Liu}, and X.~{Cao}, ``Joint
  optic disc and cup segmentation based on multi-label deep network and polar
  transformation,'' \emph{IEEE Transactions on Medical Imaging}, vol.~37,
  no.~7, pp. 1597--1605, 2018.

\bibitem{deep-supervision}
C.-Y. Lee, S.~Xie, P.~Gallagher, Z.~Zhang, and Z.~Tu, ``{Deeply-Supervised
  Nets},'' in \emph{Proc. of International Conference on Artificial
  Intelligence and Statistics}, vol.~38, 2015, pp. 562--570.

\bibitem{DiceLoss}
K.~C.~L. Wong, M.~Moradi, H.~Tang, and T.~Syeda-Mahmood, ``{3D} segmentation
  with exponential logarithmic loss for highly unbalanced object sizes,'' in
  \emph{Proc. of Medical Image Computing and Computer-Assisted Intervention
  (MICCAI)}.\hskip 1em plus 0.5em minus 0.4em\relax Springer, 2018, pp.
  612--619.

\bibitem{Two-stage}
Y.~Sun, Z.~Fu, S.~Stainton, S.~Barney, W.~Innes, J.~Hogg, and S.~S. Dlay,
  ``Automated retinal layer segmentation of {OCT} images using two-stage {FCN}
  and decision mask,'' in \emph{Proc. of IEEE International Symposium on Signal
  Processing and Information Technology}, 2019, pp. 1--6.

\bibitem{Adam}
D.~P. Kingma and J.~Ba, ``Adam: a method for stochastic optimization,'' in
  \emph{Proc. of International Conference on Learning Representations (ICLR)},
  2015, pp. 1--15.

\bibitem{seven}
G.~Mohandass, R.~A. Natarajan, and S.~Sendilvelan, ``Retinal layer segmentation
  in pathological {SD-OCT} images using boisterous obscure ratio approach and
  its limitation,'' \emph{Biomedical and Pharmacology Journal}, vol.~10, no.~3,
  pp. 1585--1591, 2017.

\end{thebibliography}
\end{document}